\documentclass[12pt]{article}

\newcommand{\fr}{\frac}
\newcommand{\lb}{\label}

\newcommand{\be}{\begin{equation}}
\newcommand{\ee}{\end{equation}}
\newcommand{\ba}{\begin{array}}
\newcommand{\ea}{\end{array}}
\newcommand{\beqa}{\begin{eqnarray}}

\newcommand{\la}{\lambda}

\newcommand{\si}{\sigma}
\newcommand{\te}{\theta}
\newcommand{\del}{\partial}
\newcommand{\eeqa}{\end{eqnarray}}
\newcommand{\ep}{\epsilon}

\newcommand{\Bc}{{\cal B}}
\newcommand{\Kc}{{\cal K}}

\newcommand{\Lc}{{\cal L}}

\newcommand{\kd}{\delta}

\begin{document}
\title{}
\author{}
\date{}
\begin{flushright}
hep-th/0302074
\end{flushright}

\vspace{1cm}

\noindent
{\large \bf Noncommutative Maxwell--Chern--Simons theory, {\mbox duality} 
and a new noncommutative Chern--Simons theory in d=3}

\vspace{1cm}
\noindent
{\footnotesize  \"{O}mer F. DAYI\footnote{E-mail: dayi@itu.edu.tr.} 

\vspace{10pt}

\noindent
{\footnotesize \it The Abdus Salam ICTP, Strada Costiera 11, 34014, Trieste--Italy}

\vspace{10pt}

\noindent
{\footnotesize \it Physics Department, Faculty of Science and
Letters, Istanbul Technical University,}

\noindent
{\footnotesize \it  80626 Maslak--Istanbul,
Turkey.}

\vspace{10pt}

\noindent
{\footnotesize \it Feza G\"{u}rsey Institute,}

\noindent
{\footnotesize \it P.O.Box 6, 81220
\c{C}engelk\"{o}y--Istanbul, Turkey. }

\vspace{2cm}

\noindent
{\bf Abstract} 

Noncommutative Maxwell--Chern--Simons theory in 3--dimensions  is
defined in terms of star product and noncommutative  fields.
Seiberg--Witten map is employed to write it in terms of
ordinary fields. 
A parent action is introduced and 
the dual action is derived.
For spatial  noncommutativity  
it is studied up to second order
in the noncommutativity parameter $\theta .$
A new noncommutative Chern--Simons action is defined in terms
of ordinary fields, inspired by the dual action. 
Moreover, a transformation between noncommuting
and ordinary fields is proposed.

\newpage

\section{Introduction}

An equivalence of
``ordinary" (commutative) and 
noncommutative gauge fields
leads to a transformation between them which is
known as
Seiberg--Witten (SW) map\cite{sw}.
This permits one to study  noncommutative gauge theories
in terms of ordinary fields. In fact,
in \cite{grs} (S) duality is incorporated into 
noncommutative Maxwell theory action
in terms of ordinary fields
after performing SW map.
In  4--dimensions if the original theory is
noncommutative Maxwell theory where
 noncommutativity is
spatial, its dual theory is a 
noncommutative gauge theory whose time variable 
is effectively noncommuting
with the other coordinates\cite{grs}.
This interesting phenomenon
is a consequence of the  fact that
the duality
transformation   includes 4--dimensional
totally antisymmetric tensor. 

In 3--dimensions the most extensively studied duality
is  between 
Maxwell--Chern--Simons (MCS)
theory and self dual theory\cite{dj}. It leads to two equivalent
descriptions of the dynamics of massive spin--1 field. 
One of its most known applications 
is bosonization in 3--dimensions\cite{fs}.
We wonder what would be the consequences of generalization
of this duality to noncommutative MCS theory.
In \cite{gho} and \cite{cm} some generalizations of
the mentioned duality to noncommutative theories
are investigated in terms of noncommuting fields.
However, duality can also be studied employing
ordinary fields in the spirit of \cite{grs}. 
Although at first sight this can appear to be
trivial due to the fact that 3--dimensional
noncommutative Chern--Simons (CS)  action becomes
the usual CS action
in terms of SW map\cite{gs}--\cite{rb},
we will show that it gives  nontrivial results.

We write
3--dimensional noncommutative MCS action  in terms
of ordinary
gauge fields utilizing  SW map. We introduce a
parent action  in terms of ordinary fields
to obtain the dual description.
We study the dual action up to the second
order in the noncommutativity parameter $\te ,$
when we let only spatial noncommutativity.
Once the dual description is obtained it  inspires 
a new noncommutative
CS theory in terms of ordinary
gauge fields. We discuss equations of motion following
from this new action. Moreover, we propose to write
it in terms of noncommuting fields as the simplest
generalization of abelian CS action. This leads to
an explicit  transformation between noncommutative  and
ordinary fields.

\section{Duality and Noncommutative MCS {\mbox Theory}}

It is  well known that noncommutativity between coordinates 
can be introduced 
in terms of
the star product
\be
\label{star}
\ast\equiv\exp \frac{i\te^{\mu\nu}}{2} \Big(
{\stackrel\leftarrow\del}_{\mu} {\stackrel\rightarrow\del}_{\nu}
-{\stackrel\leftarrow\del}_{\nu}{\stackrel\rightarrow\del}_{\mu}
\Big),
\ee
where $\te^{\mu \nu}$ are  antisymmetric constant parameters. 
Thus, one retains coordinates commuting under the usual
product.

Noncommutative MCS theory in 3--dimensions
can be defined as
\be
\lb{a1}
S=\hat{S}_M+\hat{S}_{CS},
\ee
in terms of 
the noncommutative CS  action 
\be
\lb{ncsa}
\hat{S}_{CS}=\fr{m}{2}\ep^{\mu\nu\rho} 
\int d^3x  \left( \hat{A}_\mu \del_\nu  \hat{A}_\rho +\fr{2}{3}
 \hat{A}_\mu \ast \hat{A}_\nu \ast \hat{A}_\rho \right)
\ee
and the noncommutative Maxwell theory
\be
\hat{S}_M=-\fr{1}{4}\int d^3x \hat{F}_{\mu\nu}\hat{F}^{\mu\nu}.
\ee
We employed the noncommutative field strength:
\be
\lb{fn}
\hat{F}_{\mu\nu}=\del_\mu\hat{A}_\nu-\del_\nu\hat{A}_\mu
-i\hat{A}_\mu\ast\hat{A}_\nu+i\hat{A}_\nu \ast\hat{A}_\mu .
\ee
$\hat{A}^\mu$ are not operators but they
are called noncommutative gauge fields
 in the sense that they take values
in noncommutative space.

One can show that
the noncommutative MCS action
(\ref{a1}) is invariant under the gauge transformations
\be
\hat{\kd}_{\hat{\la}}\hat{A}^\mu = \del^\mu \hat{\la}+i\hat{\la}\ast \hat{A}^\mu 
-i\hat{A}^\mu \ast \hat{\la}.
\ee

The equivalence relation between the noncommuting
$\hat{A}^\mu,\ \hat{\la}$ and the 
ordinary (commuting) gauge fields and gauge parameter
$A^\mu ,\ \la :$
\be
\lb{eqr}
\hat{A}^\mu(A) + \hat{\kd}_{\hat{\la}}\hat{A}^\mu (A) =
\hat{A}^\mu(A+\kd_\la A) ,
\ee
leads to the SW map\cite{sw}. To the first order in  $\te$
it is written explicitly as
\be
\lb{cvf}
\hat{A}^\mu=A^\mu -\te^{\rho \nu} (A_\rho\del_\nu A_\mu-\fr{1}{2}A_\rho\del_\mu A_\nu).
\ee

When the change of variables which follows from (\ref{eqr}) is 
performed
the noncommutative CS action (\ref{ncsa})
becomes the usual action\cite{gs}
\be
\lb{oocs}
\hat{S}_{CS}=\fr{m}{2}\ep^{\mu\nu\rho} 
\int d^3x   A_\mu \del_\nu  A_\rho .
\ee
Thus, in terms of the SW map the
action (\ref{a1}) can be expressed as
\be
\lb{a2}
S=\int d^3x \left\{ -\fr{1}{4}\left[ F_{\mu\nu} F^{\mu\nu} 
+{\cal L}(\te ,F) \right]
+\fr{m}{2}
\ep^{\mu\nu\rho} A_\mu \del_\nu  A_\rho \right\},
\ee
where the $\te$ dependent part can be written as
\[
{\cal L}(\te ,F) \equiv L_\te(F)+L_{\te^2}(F) +\cdots .
\]
$L_{\te^n}$  is at the nth order in $\te .$
In the following we will use only the first 
and the second order terms, which 
can be written as\cite{grs}
\beqa
L_\te (F) & = & 
2\te_{\mu \nu}F^{\nu\rho} F_{\rho \sigma} F^{\sigma \mu}
-{1}{2}\te_{\mu \nu} F^{\mu \nu} F_{\rho \sigma}F^{\rho \sigma},\lb{lte}\\
L_{\te^2} (F) & = & 
2\te_{\nu \mu} F^{\nu \rho} \te_{\rho \si } F^{\si \kd } F_{\kd \xi} F^{\xi \mu}
+\te_{\mu \nu} F^{\nu \rho} F_{\rho \si } \te^{\si \kd } F_{\kd \xi} F^{\xi \mu}
+\te_{\mu \nu} F^{\nu \mu} \te_{\rho \si} F^{\si \xi}F_{\xi \kd} F^{\kd \rho} \nonumber \\
 & & -\fr{1}{8}(\te_{\mu \nu}F^{\mu \nu})^2  F_{\rho \si} F^{\si \rho}
+\fr{1}{4}\te_{\mu \nu} F^{\nu \rho} \te_{\rho \si } F^{\si \mu } F_{\kd \xi} F^{\xi \kd }.
\eeqa

Let us introduce  the parent action
\beqa
S &  = & \int d^3x {\Big \{ } 
-\ep^{\mu\nu\rho} B_\mu \del_\nu  A_\rho  
+\fr{1}{2}B_\mu B^\mu 
+ \fr{m}{2} \ep^{\mu\nu\rho} A_\mu \del_\nu  A_\rho \nonumber \\
& &  -\fr{1}{4}\left[ L_\te(F)
+L_{\te^2}(F)+\cdots \right]  {\Big \} } . \lb{a3}
\eeqa
Equations of motion with respect to $B_\mu$ are
\[
B_\mu =\ep_{\mu \nu \rho}\del^\nu A^\rho .
\]
When we substitute $B^\mu$ with this in  (\ref{a3}) 
the noncommutative MCS action (\ref{a2}) follows.

On the other hand
the equations of motion with respect to
$A_\mu$ 
\be
\lb{eqm}
\del_\nu \left[  \ep^{\mu \nu\rho} (B_\rho - m  A_\rho ) -
\fr{1}{2}
\fr{\kd {\cal L}}{\kd F_{\nu \mu }} \right] =0,
\ee
can be solved  for $A^\mu$ as
\be
\lb{sol1}
A_\mu = \fr{1}{m}B_\mu +\fr{1}{m}b_\mu (\te, B).
\ee
We defined $b_\mu(\te , B)$ 
in terms of the equation 
\be
\lb{kb}
b_\mu (\te , B)+\fr{1}{4} \ep_{\mu \nu \rho }
{\cal K}^{\nu \rho }_\te(\fr{H}{m}+\fr{h(\te ,B )}{m} ) =0,
\ee
where $H=dB,\ h (\te ,B)=db(\te ,B)$ and
\[
\Kc_\te^{\mu \nu} (F)\equiv
\fr{\kd \Lc (\te ,F)}{\kd F_{\mu \nu}}.
\]
Obviously, $b_\mu(\te , B)$  can be expanded in powers of $\te$ as
\[
b^\mu(\te , B)=b_\te^\mu +b_{\te^2}^\mu + \cdots .
\]
When we plug the solution (\ref{sol1})
into  (\ref{a3})  the dual of noncommutative MCS action follows:
\beqa
S_D & = \int d^3x & {\Big \{ } \fr{1}{2}B_\mu B^\mu +
\fr{1}{2m}\ep_{\mu \nu \rho }\left[ B^\mu \del^\nu B^\rho
+b^\mu(\te ,B) \del^\nu b^\rho(\te ,B) \right] \nonumber \\
& & -\fr{1}{4}\Lc \left(\te ,\fr{H}{m}+\fr{h(\te ,B)}{m}\right) { \Big\} }.
\lb{dac}
\eeqa

In the most general case we can add
to the solution (\ref{sol1}) the term
$\del_\mu \kappa ,$ where $\kappa$
is an arbitrary function. 
However, this alters
the dual action (\ref{dac}) only up to a surface term
which we can drop. Obviously, this is equivalent
to choose a vanishing $\kappa .$

When the noncommutativity is along the spatial coordinates:
\be
\lb{spn}
\te^{ij}=\te \ep^{ij},\ \te^{0i}=0,
\ee
the first order term (\ref{lte})
can be written as
\be
\lb{fol}
L_\te(F)=\te F_{12}F_{\mu \nu}F^{\mu \nu}.
\ee
Now, we can solve (\ref{kb}) to obtain 
\beqa
b_\te^0 & = & \fr{\te}{m^2}\left[ H_{\mu \nu}H^{\mu \nu} +
2H_{12}H_{12}\right] , \nonumber \\
b_\te^1 &  = & \fr{2\te }{m^2} H_{12}H_{02}, \lb{bt} \\
b_\te^2 & = & \fr{2\te }{m^2} H_{12}H_{10}. \nonumber
\eeqa
When we use these  in (\ref{dac}) explicit form of the 
dual action
to the second order in $\te$ follows.  
To the second order in $\te$ (\ref{dac}) can be written as
\begin{eqnarray*}
S_{D,(2)} & \equiv \int d^3x & {\Big \{ } \fr{1}{2}B_\mu B^\mu +
\fr{1}{2m}\ep_{\mu \nu \rho } B^\mu \del^\nu B^\rho
-\fr{1}{4}L_\te \left(\fr{H}{m}\right) 
+\fr{1}{2m}\ep_{\mu \nu \rho }b^\mu_\te \del^\nu b^\rho_\te \\ 
& &  -\fr{1}{4}L_{\te^2} \left(\fr{H}{m}\right) 
+\fr{2}{m^2}\te_{\mu \nu}\left(
H^{\nu\rho} h_{\te \rho \sigma} H^{\sigma \mu}
+2H^{\nu\rho} H_{\rho \sigma} h_\te^{\sigma \mu}
\right) \\
& & -\fr{12}{m^2}\te_{\mu \nu}\left(
 h_\te^{\mu \nu} H_{\rho \sigma}H^{\rho \sigma}
+2 H^{\mu \nu} h_{\te \rho \sigma}H^{\rho \sigma}
\right)
{ \Big\} },
\end{eqnarray*}
where $h_\te^{\mu \nu}=\del^\mu b_\te^\nu -\del^\nu b_\te^\mu .$

Although, the dual actions (\ref{a2}) and (\ref{dac}) are obtained from
the parent action (\ref{a3}), it is not guaranteed 
that they yield the same partition function. 
Indeed, we deal only with the classical aspects.
Quantum corrections may oblige us to regulate
the action (\ref{a3}) \cite{grs}, which is not addressed 
in this  work.

\section{ A New Noncommutative CS Theory}

When SW map is employed the
noncommutative CS action (\ref{ncsa})
becomes the ordinary CS action (\ref{oocs}) \cite{gs}.
Thus, a noncommutative CS theory formulated  in terms
of the ordinary gauge fields $A_\mu$ and the
noncommutativity parameter $\te$
is not available.
However, we can utilize the action
$S_D$ (\ref{dac}) to define a new noncommutative
abelian CS theory in terms of the ordinary 
gauge fields $B^\mu $ and the noncommutativity
parameter  $\te_{\mu \nu } :$

Setting $\te^{\mu \nu}=0$ 
and dropping the ordinary mass term  $B^2$ 
in $S_D$ (\ref{dac})  lead to the
ordinary abelian  CS action:
\be
\lb{csbb}
S_{CS}[B] =\fr{M }{2}\int d^3x\ 
\ep_{\mu \nu \rho } B^\mu \del^\nu B^\rho ,
\ee
where $M \equiv 1/m .$ We would like to take advantage
of this observation to
define a new noncommutative
abelian CS theory in terms of the ordinary 
gauge fields $B^\mu $ as
\be
\lb{dcs}
S_{{\rm NCS}}  =   \int d^3x \left\{
\fr{M}{2} \ep_{\mu \nu \rho }\left[ B^\mu \del^\nu B^\rho
+b^\mu(\te ,B) \del^\nu b^\rho(\te ,B) \right] 
 -\fr{1}{4}\Lc \left (\te ,MH+Mh(\te ,B)\right)
\right\} ,
\ee
by dropping the $B^2$ term in (\ref{dac}).
Obviously, this action is invariant under the abelian
 gauge transformations $\kd B_\mu=\del_\mu \lambda$
and yields the ordinary 
CS theory (\ref{csbb}) when one sets  $\te=0.$

Equations of motion are
\be
\lb{deqm}
\ep_{\mu \nu \rho} \del^\nu \left(  B^\rho 
+ b^\rho (\te ,B))  \right)
-4\ep_{\si \nu  \rho}\del^\kappa 
\left[ G^\si _{\kappa \mu}(H) \del^\nu b^\rho \right] =0,
\ee
where we defined
\be
\lb{dg}
\frac{\kd b^\mu (\te ,B(y) )}{\kd H^{\nu \rho }(x)  }
= G_{\nu \rho }^\mu (H) \kd^3 (x-y).
\ee
Observe that the simplest solution of (\ref{deqm}) is 
\[
H^{\mu \nu }=0,
\]
which is independent of $\te .$ To the first order in
$\te$ the equations of motion (\ref{deqm}) 
get the simple form
\be
\ep_{\mu \nu \rho} \del^\nu (B^\rho 
+b^\rho_\te )  =0.
\ee

The SW map (\ref{eqr}) expresses the noncommutative gauge fields
$\hat{A}_\mu$ in terms of the ordinary gauge fields $A_\mu$
utilizing the equivalence relation (\ref{cvf}). 
A transformation between noncommutative and ordinary fields 
can also be derived by assuming an equivalence relation between
the action (\ref{dcs}) and another one written by introducing
some fields $\Bc (B, \te )$ taking values 
in noncommutative space. However, there is no unique choice
for the latter action.
One should make an assumption about the  
form of the action in terms 
of the noncommuting fields $\Bc (B, \te ).$
Let us suppose that the action in terms of 
the noncommutative fields $\Bc (B, \te )$, 
is in the same form as the abelian
CS theory:
\be
\lb{csin}
S_{NCS}\equiv \fr{M}{2}\int d^3x
\ep_{\mu \nu \rho}\Bc^\mu  (B, \te) \del^\nu \Bc^\rho (B, \te ).
\ee
One can show that there exists a transformation between
$\Bc^\mu  (B, \te)$ and $B^\mu .$
Indeed, one can solve for $\Bc$ perturbatively in $\te$. 
At the first order in $\te$ one should solve
\be
\ep_{\mu \nu \rho}B^\mu_\te H^{\rho \nu} 
=L_\te (M H),
\ee
where $B_\te^\mu =\del \Bc^\mu /\del \te|_{\te =0}.$
There is not a unique solution. For instance,
when the noncommutativity is only spatial (\ref{spn}), a solution is
\[
B_\te^\mu = \frac{M\te}{2}  H_{21}\ep^{\mu \nu \rho}H_{\nu \rho} .
\]

Although the assumption  (\ref{csin})
is very plausible, in principle one may
define some other actions in terms of fields taking values 
in noncommutative space.
Nevertheless, the assumed form of the action (\ref{csin}) is
shown to yield a map between the noncommutative gauge 
fields $\Bc^\mu  (B, \te)$ and the ordinary ones $B^\mu$ 
which is not the SW map (\ref{cvf}). Moreover,
the form of the action (\ref{csin}) can be useful to
generalize this construction to nonabelian gauge theories.

\vspace{.5cm}

\noindent
{\bf Acknowledgment:} I would like to thank referee for useful
comments.

\end{document}